\documentclass[fleqn,usenatbib]{mnras}

\usepackage[T1]{fontenc}
\usepackage{ae,aecompl}

\usepackage{graphicx} 
\usepackage{amsmath}	
\usepackage{amssymb}	
\usepackage{amsfonts}
\usepackage{hyperref}
\usepackage{newtxtext,newtxmath}
\usepackage[normalem]{ulem}

\title[Uniformity from dynamical sculpting]
{Intra-system uniformity: a natural outcome of dynamical sculpting}

\author[Lammers et al.]
{Caleb Lammers,$^{1, 2}$
Sam Hadden$^{1}$,
Norman Murray$^{1, 2}$
\\
$^{1}$Canadian Institute for Theoretical Astrophysics, University of Toronto, 60 St.\ George Street, Toronto, ON M5S 3H8, Canada\\
$^{2}$Department of Physics, University of Toronto, 60 St.\ George Street, Toronto, ON M5S 1A7, Canada\\
}

\pubyear{2023}

\begin{document}
\maketitle

\begin{abstract}

\noindent There is evidence that exoplanet systems display intra-system uniformity in mass, radius, and orbital spacing (like ``peas in a pod'') when compared with the system-to-system variations of planetary systems. This has been interpreted as the outcome of the early stages of planet formation, indicative of a picture in which planets form at characteristic mass scales with uniform separations. In this paper, we argue instead that intra-system uniformity in planet sizes and orbital spacings likely arose from the dynamical sculpting of initially-overly-packed planetary systems (in other words, the giant impact phase). With a suite of $N$-body simulations, we demonstrate that systems with random initial masses and compact planet spacings naturally develop intra-system uniformity, in quantitative agreement with observations, due to collisions between planets. Our results suggest that the pre-giant impact planet mass distribution is fairly wide and provide evidence for the prevalence of dynamical sculpting in shaping the observed population of exoplanets.

\end{abstract}

\begin{keywords}
celestial mechanics -- planets and satellites: dynamical evolution and stability -- planets and satellites: formation
\end{keywords}

\section{Introduction}
\label{introduction}

One of the most striking findings of the {\it Kepler} mission, and later surveys, is the staggering abundance of compact multiplanet systems \citep[e.g.,][]{Borucki2011, Fabrycky2014, Udry2019}. These systems are characterized by multiple close-in planets, typically with masses a few times that of the Earth, low inclinations, and low eccentricities \citep[e.g.,][]{Lissauer2011, Figueira2012, Fressin2013, VanEylen&Albrecht2015, Hadden&Lithwick2017}. It was first pointed out in \citet{Lissauer2011} that planets in compact multiplanet systems possess similar radii to their neighbours. Later, \citet{Weiss2018} confirmed this finding with a larger sample of multiplanet systems and found additional consistency in the orbital spacings of planets, dubbing this propensity for intra-system uniformity the ``peas-in-a-pod'' trend. Subsequent studies have also found a significant degree of intra-system uniformity in planet masses, among systems for which mass measurements are available \citep{Millholland2017, Wang2017, Goyal&Wang2022, Otegi2022}. Although biases introduced by selection effects complicate the interpretation of the observed intra-system uniformity trends \citep{MurchikovaTremaine2020, Zhu2020}, there is evidence that the observed trends are, at least partially, astrophysical in origin \citep{He2019, WeissPetigura2020, Otegi2022, Weiss2022}. Thus, finding astrophysical mechanisms that can produce intra-system uniform planet sizes and spacings is a crucial part of understanding the formation and evolution of compact multiplanet systems.

Theoretical explanations for the peas-in-a-pod trend have principally focused on the early stages of planet assembly. \citet{Adams2019} and \citet{Adams2020} demonstrated that any planet assembly process in which dissipation drives the system to its minimum-energy state (subject to mass, angular momentum, and orbital spacing constraints) will produce nearly equal-mass planets, provided planets are smaller than $\lesssim\,40\,M_{\oplus}$. \citet{Batygin&Morbidelli2023} presented a super-Earth formation model in which isolation and orbital migration regulate mass growth, setting a characteristic planet mass scale that varies from system to system. \citet{Mishra2021} similarly predicts that the peas-in-a-pod trend arises at early times, before the dispersal of the protoplanetary disk, using a population synthesis model to simulate the formation of planetary systems.

In this paper, we explore the consequences of post-formation dynamical evolution on the degree of mass and spacing uniformity in compact systems of small planets. Typical observed transiting exoplanet systems, with innermost orbital periods on the order of ${\sim}\,10$\,days and ages of ${\sim}\,1\,-\,10$\,Gyr, are separated from the planet assembly stage by $\sim 10^{10}$\,--\,$10^{12}$ orbits. $N$-body simulations of compact multiplanet systems show that dynamical instabilities frequently arise over long timescales in such systems \citep{Chambers1996, Smith&Lissauer2009, Obertas2017}. Indeed, the spacings of observed exoplanet systems suggest that such instabilities have, in fact, played a prominent role in sculpting the population observed today \citep{Pu&Wu2015}. Intra-system uniformity has been observed in simulations of super-Earth/sub-Neptune formation \citep{MacDonald2020, Izidoro2022}, but it is difficult to deduce the influence of giant impacts on the degree of uniformity in such studies. To this end, \citet{Goldberg&Batygin2022} studied the effects of dynamical sculpting on equally-spaced (resonant chain) systems initialized with nearly-identical planet masses. They found that dynamical evolution does not degrade the degree of mass/spacing uniformity too significantly, in that the outcomes of their simulations are broadly consistent with the degree of uniformity of observed systems. Here we demonstrate that, strikingly, the dynamical sculpting of systems initialized with random masses and random spacings naturally leads to intra-system uniformity, in quantitative agreement with the observed population of exoplanets.

\section{Observations of intra-system uniformity}
\label{multiplanetsys}

\begin{figure}
\centering
\includegraphics[width=0.4\textwidth]{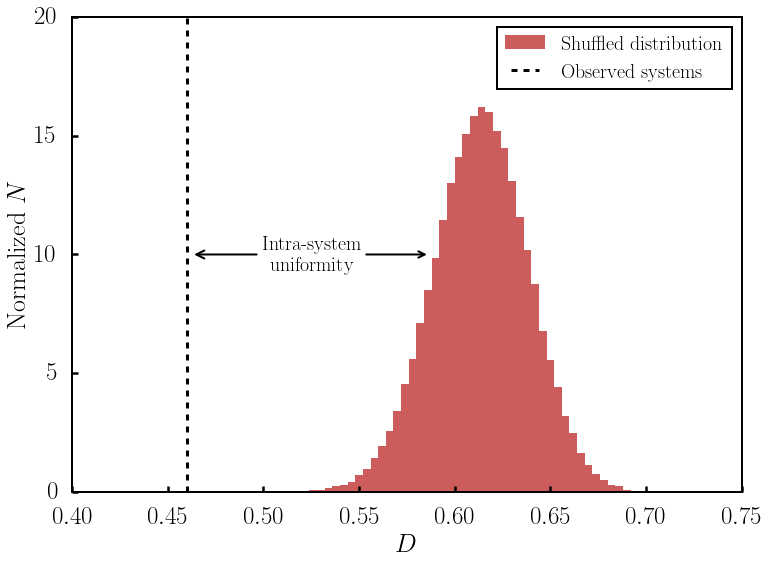}
\caption{Comparison of the mean mass dispersion $D$ between 100 observed planetary systems and $100$,$000$ Monte Carlo realizations in which we shuffled the planets among the systems. The observed $D_{\mathrm{obs}}\,{=}\,0.46$ lies well below the distribution of shuffled $D$ values (significance $S_{\mathrm{obs}}\,{=}\,6.2$), indicating that compact multiplanet systems exhibit intra-system uniformity in their masses.}
\label{obs_uniformity}
\end{figure}

Previous studies have defined various quantitative metrics to characterize the degree of mass and radius uniformity. We adopt the mean normalized mass dispersion,
\begin{equation} 
\label{D}
D\,{=}\,\frac{1}{N_{\mathrm{sys}}} \sum_{i=1}^{N_{\mathrm{sys}}}\frac{\sigma_{m_i}}{\overline{m}_i}\,{=}\,\frac{1}{N_{\mathrm{sys}}} \sum_{i=1}^{N_{\mathrm{sys}}}\sqrt{\frac{\sum_{j=1}^{N_{\mathrm{pl}_i}} (m_j - \overline{m}_i)^2}{\overline{m}_i^2 (N_{\mathrm{pl}}-1)}}~,
\end{equation}
as defined in \citet{Goldberg&Batygin2022}, where $N_{\mathrm{pl}_i}$, $\overline{m}_i$, and $\sigma_{m_i}$ are the number of planets, mean mass, and mass standard deviation of the $i$th system, respectively. We have confirmed that our conclusions do not depend on the particular choice of metric. We calculate $D$ for the observed population of small planets using data from the NASA Exoplanet Archive.\footnote{\url{exoplanetarchive.ipac.caltech.edu} (accessed 7 November 2022)} We include in our sample all confirmed multiplanet systems for which at least two planet masses have been measured and all planets are smaller than $30\,M_\oplus$ (mass intra-system uniformity is not seen in systems with known massive planets; \citealt{Wang2017}). For the 100 confirmed exoplanet systems consisting of 266 planets that satisfy our selection criteria, we find $D_\mathrm{obs}\,{=}\,0.46$.

As in previous studies \citep[e.g.,][]{Millholland2017, Wang2017, Otegi2022}, we assess the significance of the intra-system uniformity in our sample by comparing its $D$ value to a Monte Carlo distribution of values generated by randomly shuffling planets among the set of systems. Figure~\ref{obs_uniformity} compares our sample's value of $D$ to the distribution of $100$,$000$ $D$ values generated by repeatedly shuffling the planets and re-calculating $D$. The value of
\begin{equation} 
\label{S}
S\,{=}\,\frac{\overline{D}_{\mathrm{shuffled}}\,-\,D}{\sigma_{D_\mathrm{shuffled}}},
\end{equation}
where $\overline{D}_{\mathrm{shuffled}}$ and $\sigma_{{D}_\mathrm{shuffled}}$ are, respectively, the mean and standard deviation of the shuffled $D$ values, provides a useful metric for the significance of the intra-system uniformity trend (as compared with \emph{inter}-system diversity). We compute a value $S_{\mathrm{obs}}\,{=}\,6.2$ for our sample of multiplanet systems. We caution that complex selection effects may bias the observed values of $D$ and $S$ with respect to the underlying true population of systems of small exoplanets. Nonetheless, the values of $D$ and $S$ computed for our sample provide useful points of comparison for the results of the $N$-body simulations presented below. When we repeat our calculations for the sample used in \citet{Millholland2017}, we find comparable values of $D_{\mathrm{obs}}\,{=}\,0.43$ and $S_{\mathrm{obs}}\,{=}\,7.5$.

\section{$N$-body simulations}
\label{Nbody}

\subsection{Simulation setups}
\label{simsetup}

We study systems of ten initially-circular, coplanar\footnote{The results of our fiducial simulation ensemble are comparable ($D_{\mathrm{final}}\,{\sim}\,0.5$, $S_{\mathrm{final}}\,{\sim}\,5$) if the planets begin instead with small initial inclinations (${\sim}\,1^\circ$). Note, however, that the final distribution of inclinations in this case is somewhat wider than that of observed systems ($\sigma_i\,{\sim}\,2.5^\circ$). See also \citet{Ghosh&Chatterjee2023}, which came to similar conclusions with a different set of initial conditions, including non-zero inclinations.} planets orbiting a solar-mass star. We place the innermost planet at $a_1\,{=}\,0.1$\,AU and randomly space the remaining nine planets by drawing the period ratios of each successive adjacent planet pair, $P_{i+1}/P_i$, uniformly from $[1.10,\,1.50)$. Initial longitudes are drawn uniformly from the range $[0,\,2\pi)$ and initial planet masses are drawn from a normal distribution with mean $\mu\,{=}\,3$\,M$_{\oplus}$ and standard deviation $s_m$. Any negative masses are rejected and re-drawn. It is straightforward to show that, for this mass distribution, the expected value of $D$ depends on the single parameter $s_m/\mu$ and ranges from $D\,{=}\,0$ for $s_m/\mu\,{=}\,0$ to $D\,{=}\,\sqrt{{\pi/2-1}}\,{\approx}\,0.76$ for $s_m/\mu\,{\rightarrow}\,\infty$. We set planet radii using the mass-radius relationship $(M/M_\oplus)\,{=}\,2.7\,(R/R_\oplus)^{1.3}$ from \citet{Wolfgang2016}. The initial conditions of our simulations (i.e., closely-spaced planets on low-eccentricity orbits) were chosen to reflect the expected state of planetary systems at the time the gas disk disperses.

$N$-body simulations were carried out with the {\sc MERCURIUS} hybrid integrator \citep{Rein2019} from the {\sc REBOUND} open-source code \citep{Rein&Liu2012}. This integration scheme relies on a high-order adaptive integrator ({\sc IAS15}; \citealt{Rein&Spiegel2015}) during close encounters, and otherwise uses a standard fixed-timestep Wisdom-Holman integrator ({\sc WHFast}; \citealt{Wisdom&Holman1991, Rein&Tamayo2015}). We adopt a timestep of $P_1/20$ for the WHFast integrator and integrate all systems for $10^9\,P_1$ (${\sim}\,30$\,Myr for $a_1 = 0.1~\mathrm{AU}$). Planet collisions are treated as perfect inelastic collisions in which the total mass, momentum, and volume are conserved. Previous work indicates that this is a reasonable assumption for compact systems of small planets \citep{Poon2020, Esteves2022, Goldberg&Batygin2022}. Below, we discuss the emergence of intra-system uniformity due to dynamical sculpting.

\begin{figure*}
\centering
\includegraphics[width=0.80\textwidth]{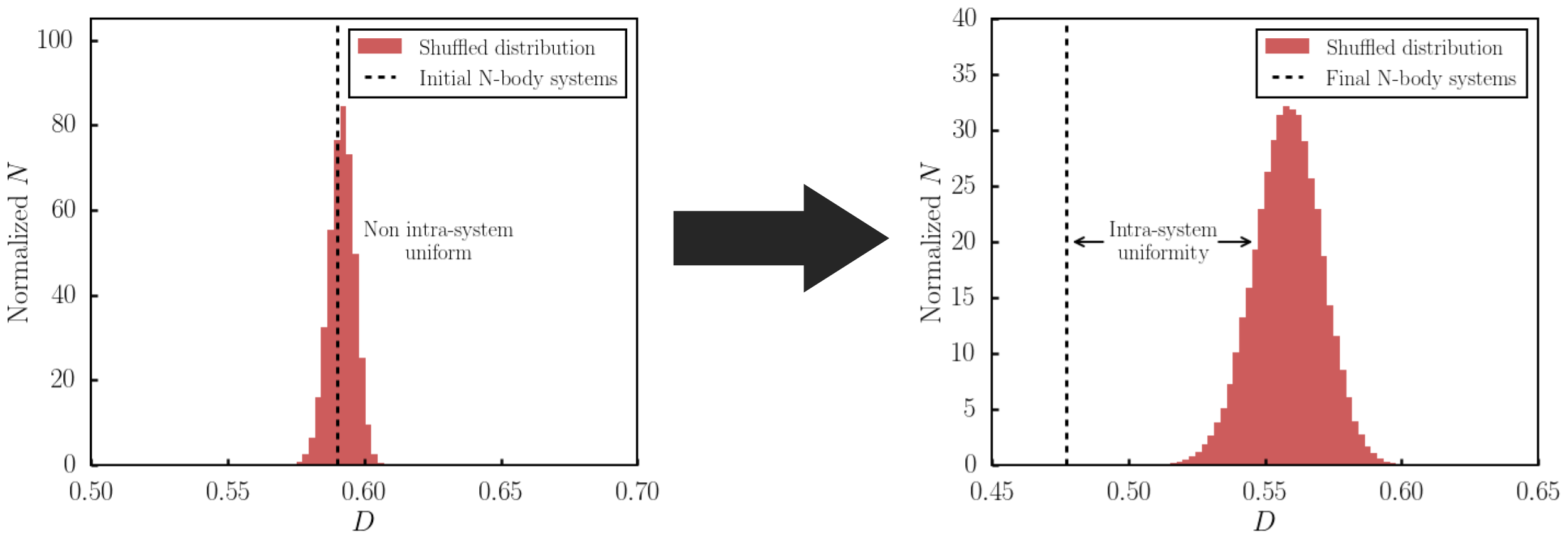}
\caption{Same plot as Fig.~\ref{obs_uniformity}, shown now for the initial and final states of $100$ simulated planetary systems. These systems were initialized with $D_{\mathrm{initial}}\,{=}\,0.59$ and $S\approx 0 $ (left panel). However, collisions over $10^9\,P_1$ of dynamical evolution cause the simulated systems to develop intra-system mass uniformity with $D_\mathrm{final}\,{=}\,0.48$ and $S_{\mathrm{final}}\,{=}\,6.6$ (right panel).}
\label{fig:sim_uniformity}
\end{figure*}

\subsection{Emergence of intra-system mass uniformity}
\label{nbodyresults1}

\begin{figure}
\centering
\includegraphics[width=0.4\textwidth]{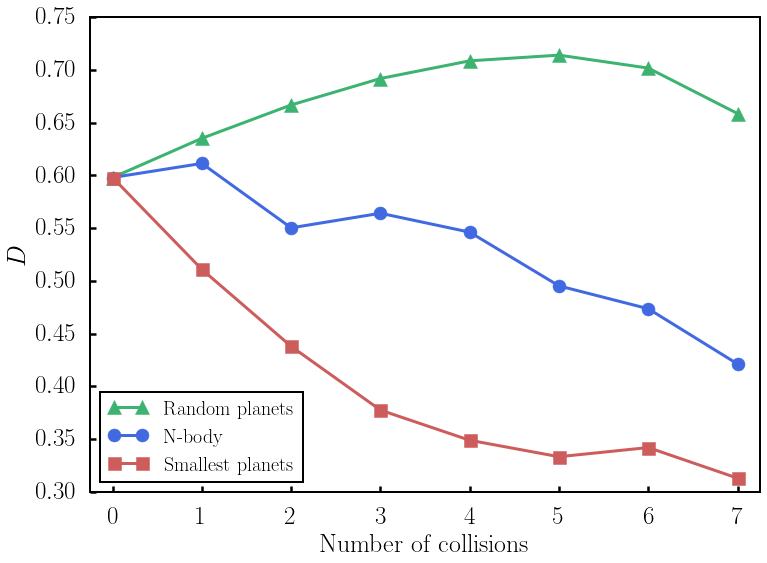}
\caption{Blue circles show the evolution of the mean mass dispersion, $D$, with the number of collisions for the systems in our fiducial $N$-body ensemble. Green triangles show the results of evolving the initial ensemble of systems by iteratively selecting random adjacent planet pairs to collide. Red squares show the results of selecting the smallest planet to collide with its smallest neighbour at each iteration. The $N$-body results reside between the ``random planets'' and ``smallest planets'' models, demonstrating that collisions typically, but not always, occur between smaller planets, thereby lowering $D$.}
\label{fig:Dtest}
\end{figure}

Our fiducial simulation ensemble consists of $100$ systems initialized according to the procedure above with a mass standard deviation of $s_m\,{=}\,3.0$\,M$_{\oplus}$. Figure~\ref{fig:sim_uniformity} summarizes the emergence of intra-system mass uniformity in this ensemble due to dynamical evolution: the left-hand panel shows that the initial planet mass distribution gives $D_{\mathrm{initial}}\,{=}\,0.59$ and $S_{\mathrm{initial}}\,{\approx}\,0$, since planet masses are all drawn from a single global distribution. The right-hand panel shows that, after $10^9\,P_1$ of dynamical evolution, the ensemble obtains $D_\mathrm{final}\,{=}\,0.48$ in close agreement with the observed value, $D_{\mathrm{obs}}\,{=}\,0.46$. When compared with shuffled configurations, the final synthetic systems possess intra-system uniformity with significance $S_{\mathrm{final}}\,{=}\,6.6$, similar to that of observed systems ($S_\mathrm{obs}\,{=}\,6.2$).

It is not obvious that collisions between planets should drive systems towards mass uniformity (i.e., lower $D$). In fact, when systems begin with sufficiently uniform masses, collisions inevitably degrade the uniformity and increase $D$ (see \citealt{Goldberg&Batygin2022}). The evolution of $D$ is more subtle when systems are initialized with a spread of planet masses. If collisions in our ensemble merely occurred randomly between adjacent planet pairs, the value of $D$ actually \emph{increases} above its initial value. This is illustrated in Fig.~\ref{fig:Dtest}, which compares the evolution of $D$ due to collisions in our $N$-body ensemble with two simple experiments. In the first experiment, we start with the set of initial planetary systems in our fiducial ensemble and then select adjacent pairs at random to combine. With random collisions, $D$ increases with the number of collisions until $5$ or fewer planets remain, and always remains well above the $N$-body ensemble's final value of $D_\mathrm{final}\,{=}\,0.48$. The tendency for dynamical evolution to increase the ensemble's degree of intra-system uniformity is due to the fact that collisions are more likely to involve lower-mass planets. In particular, collisions in the $N$-body simulations shown in Fig.~\ref{fig:sim_uniformity}, involve the smallest planet 47\% of the time, whereas just 18\% involve the largest planet (and ${\sim}\,90\%$ of collisions occur between adjacent planets). 

For comparison, the second experiment plotted in Fig.~\ref{fig:Dtest} shows the result of evolving the initial ensemble by combining the smallest surviving planet with its smallest neighbour during each collision. Exclusively selecting the smallest planet for collisions significantly increases the degree of uniformity; in fact, the $D$ value of this ensemble differs by no more than 5\% from the minimum possible $D$ that can be achieved by combining adjacent pairs at each step. The propensity for collisions to preferentially involve the lowest-mass planets is unsurprising, given that such planets typically require the smallest transfers of orbital angular momentum to cross orbits with their neighbours.

\begin{figure*}
\centering
\includegraphics[width=0.6\textwidth]{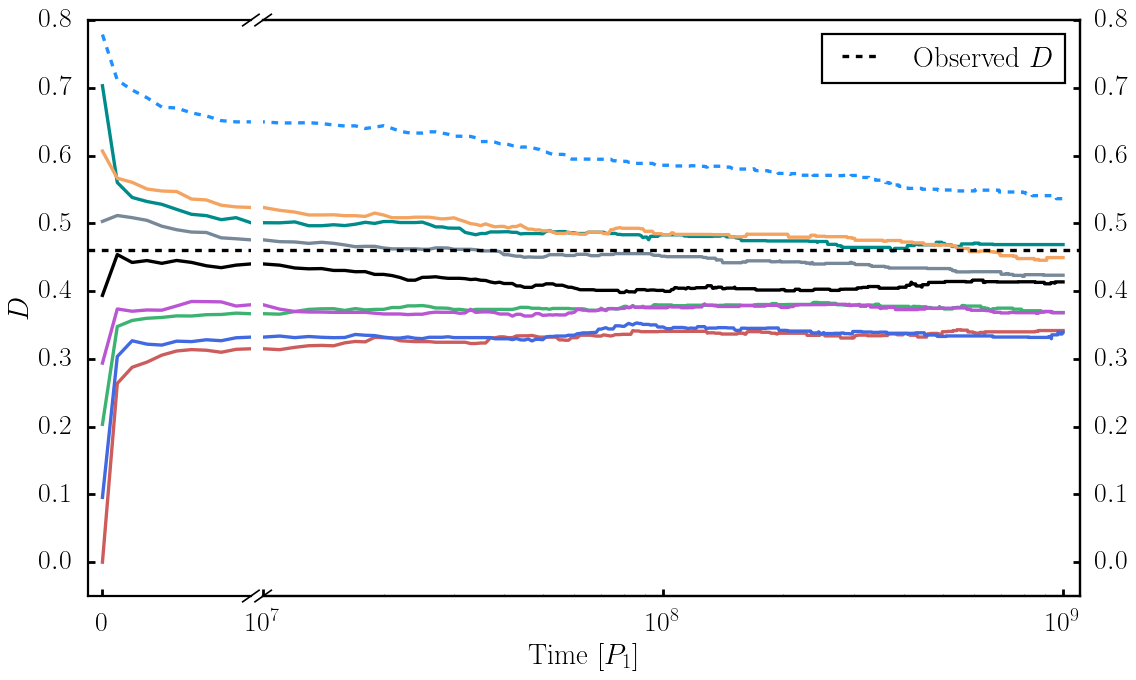}
\caption{Evolution of the mean mass dispersion, $D$, over $10^9\,P_1$ for simulations in which $D_{\mathrm{initial}}$ ranges from $0.0$\,--\,$0.8$. Planet masses are normally-drawn (solid lines) in all simulations except one, where masses are drawn according to a power law (dotted line). Time is plotted linearly until $10^7\,P_1$, after which time is shown logarithmically. Collisions between planets naturally degrade the mass uniformity of initially-uniform systems, and conversely, drive systems with a spread of initial masses towards uniformity because collisions tend to occur among smaller planets, resulting in $D_{\mathrm{final}}\,{=}\,0.3\,-\,0.5$. However, the spread of initial masses is not forgotten, and the observed value of $D_{\mathrm{obs}}\,{=}\,0.46$ favours a larger value of $D_{\mathrm{initial}}$ (and potentially a non-normal initial mass distribution).}
\label{fig:trackingD}
\end{figure*}

Next, we explore the effects of varying the initial planet mass distribution on the evolution of intra-system uniformity. We simulate $8$ ensembles of $100$ planetary systems, varying $s_m$ so that $D_\mathrm{initial}$ ranges from $0.0\,{-}\,0.7$. We also simulate an ensemble with initial planet masses drawn from the range $0.1$\,--\,$10\,M_{\oplus}$ according to the power law $dN/dm\,\propto\,m^{-1/2}$, resulting in $D_\mathrm{initial}=0.78$.

Figure~\ref{fig:trackingD} shows the time evolution of $D$ for each of our simulated ensembles. In ensembles with initially-similar planet masses ($D_\mathrm{initial}\,\lesssim\,0.4$), collisions cause $D$ to increase, whereas initially-diverse systems become more uniform. As in our fiducial ensemble, we find that the increasing intra-system uniformity in these systems is driven by collisions preferentially involving systems' smallest planets. In fact, even in ensembles where $D$ increases, smaller planets are more likely to be involved in collisions; however, because the masses of the smallest planets do not differ substantially from the most massive planets, the resulting post-collision systems have a higher $D$. As a result, dynamical evolution leads $D$ to be attracted towards the range $0.3$\,--\,$0.5$, irrespective of $D_{\mathrm{initial}}$. That said, it is apparent from Fig.~\ref{fig:trackingD} that the ensembles retain some memory of their $D_{\mathrm{initial}}$: initially-uniform masses ($D_{\mathrm{initial}}\,{\approx}\,0$) fail to climb to the relatively-high $D_{\mathrm{obs}}\,{=}\,0.46$ of observed systems, and their evolution appears to have largely halted before $10^9P_1$. The observed value of $D$ therefore favours initially-non-uniform masses ($D_{\mathrm{initial}}\,{\gtrsim}\,0.5$). In this way, $D$ provides a useful tool for probing the initial (pre-giant impact phase) planet mass distribution, and our results suggest a wide (and potentially non-normal) initial mass distribution, rather than a characteristic initial mass.

We find a final value of $S\,{=}\,6.6$ for our nominal $N$-body ensemble and a range spanning $S\,{\approx}\,5$\,--\,$10$ for the other ensembles. The value of $S$ reflects both the degree of uniformity among individual systems as well as the level of \emph{inter}-system diversity in total planet mass and planet multiplicity. Our simulations therefore demonstrate that the dynamical evolution of a population of planetary systems in which initial planet masses are independent of the particular host system can produce both intra-system uniformity (i.e., small $D$) and inter-system diversity (i.e., large $S$). Inter-system diversity arises in our simulated ensembles due to conservation of mass ($\sum_{i\,{=}\,1}^N m_i\,{\approx}\,20$\,--\,$60$\,M$_{\oplus}$), which causes planets in systems with fewer surviving planets to generally be more massive than planets in systems with more planets. A prediction of this scenario is a negative relationship between planet multiplicity and planet mass, which we find to be tentatively supported in our sample of observed systems. Adding inherent system-to-system variation to the total planet masses increases the value of $S_{\mathrm{initial}}$ and $S_{\mathrm{final}}$, but as illustrated by Fig.~\ref{fig:sim_uniformity}, no initial system-to-system variation is required to reproduce the value of $S_{\mathrm{obs}}$.

\subsection{Emergence of orbital spacing uniformity}
\label{nbodyresults2}

The above discussion has focused on the intra-system uniformity of planet masses. However, exoplanet systems also display intra-system uniformity in the orbital spacing of their planets, which is typically quantified by the correlation between the spacing of inner and outer planet pairs ($P_{i+2}/P_{i+1}$ vs.\ $P_{i+1}/P_{i}$; \citealt{Weiss2018}). Recently, \citet{Goldberg&Batygin2022} argued that uniformity in orbital spacing arises due to the dynamical sculpting of planetary systems that begin in resonant chains with nearly-identical masses (i.e., low $D_{\mathrm{initial}}$).

We find that neither resonant chains nor initially-similar masses are required for dynamical sculpting to produce this trend. In our randomly-spaced systems initialized with a spread of planet masses, a significant correlation between the spacing of outer and inner planet pairs emerges early in the $N$-body simulations (${\sim}\,10^6\,P_1$). At these early times, collisions occur primarily between tightly-packed planet pairs ($P_{i+1}/P_i\,{\lesssim}\,1.25$), driving the systems towards uniform planet spacings. In this way, orbital spacing uniformity arises naturally as a result of the dynamical relaxation of our simulated systems, in which collisions tend to occur between closely-spaced planets. The correlation between spacings of adjacent planet pairs in our final synthetic systems (Pearson-$R=\,0.39$, $p$-value$\,=2\,\times\,10^{-10}$) quantitatively agrees with that of the observed systems from \citet{Weiss2018} (Pearson-$R\,=\,0.42$, $p$-value$\,=2\,\times\,10^{-8}$).

\subsection{Orbital architectures}
\label{obssimcomp}

\begin{figure*}
\centering
\includegraphics[width=0.8\textwidth]{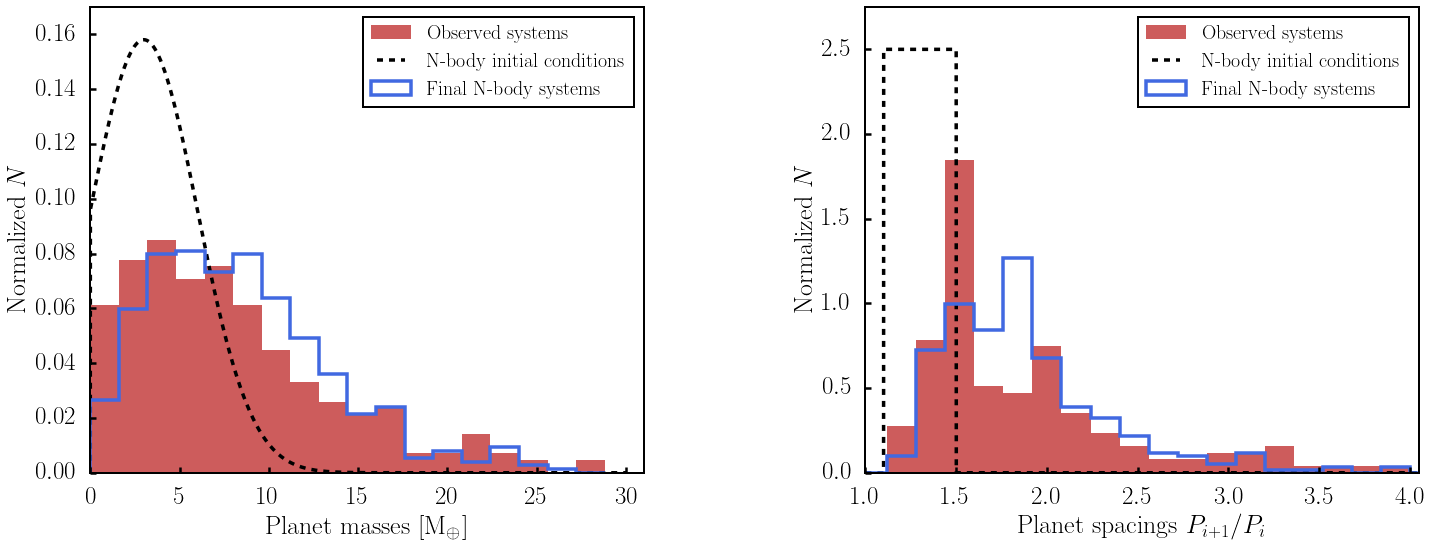}
\caption{Histograms of planet masses (left) and adjacent planet period ratios (right) for the observed systems in Fig.~\ref{obs_uniformity} and our simulated systems in Fig.~\ref{fig:sim_uniformity}. Despite the simplicity of the simulation initial conditions (normally-drawn planet masses and uniformly-drawn period ratios), dynamical evolution results in planet masses and orbital spacings that broadly agree with that of observed compact multiplanet systems.}
\label{comparison}
\end{figure*}

Finally, we can compare the orbital architectures of our fiducial simulated systems from Fig.~\ref{fig:sim_uniformity} to observed systems. Dynamical instabilities reduce the planet multiplicity of the synthetic systems from $10$ to a mean of $4.7$, in agreement with the inferred mean multiplicity of {\it Kepler\/} systems \citep{Traub2016, Zink2019}. Figure~\ref{comparison} compares the distribution of planet masses and adjacent-planet period ratios between observed systems and our simulated systems. Over $10^9\,P_1$, dynamical sculpting transforms the initially-compact, ten-planet systems into systems with planet masses and spacings that are broadly consistent with observed compact multiplanet systems. The final eccentricities of our simulated systems have a best-fit Rayleigh scale parameter $\sigma_e\,{=}\,0.03$, slightly lower than that found in previous simulations \citep{Hansen&Murray2013, Izidoro2017, Goldberg&Batygin2022} but still above that of transit timing variation (TTV) systems \citep{Hadden&Lithwick2014, Hadden&Lithwick2017}. The eccentricities of many TTV systems lie well below the boundary for stability, indicating that some subsequent eccentricity damping may be required \citep{Yee2021}.

\section{Discussion}
\label{discussion}

The finding that exoplanet systems exhibit significant intra-system uniformity in mass, radius, and orbital spacing was an unanticipated development of recent exoplanet surveys. This has been interpreted as a consequence of the early planet formation process, and several theoretical explanations for the emergence of intra-system uniformity have been put forward. In this work, we have demonstrated that intra-system uniformity in planet masses and planet spacings can arise naturally from the dynamical sculpting of compact multiplanet systems. Specifically, our $N$-body simulations demonstrate that collisions typically occur between small planets, driving systems towards intra-system mass uniformity (mass dispersion $D_{\mathrm{final}}\,{=}\,0.48$ and significance $S_{\mathrm{final}}\,{=}\,6.6$) that agrees quantitatively with that of observed systems ($D_{\mathrm{obs}}\,{=}\,0.46$ and $S_{\mathrm{obs}}\,{=}\,6.2$). Collisions, particularly at early times, also tend to occur between closely-spaced planets, resulting in uniform planet spacings (Pearson-$R$\,$=\,0.39$) that are similar to that of observed systems (Pearson-$R$\,$=\,0.42$). Furthermore, interestingly, we find that a wide spread of initial planet masses ($D_{\mathrm{initial}}\,{\gtrsim}\,0.5$), rather than a characteristic initial planet mass ($D_{\mathrm{initial}}\,{\approx}\,0.0$), is required to reproduce the extent of the mass uniformity of observed systems after the giant impact phase.

{\it Kepler\/} systems have been found to be ``dynamically-packed,'' in that additional planets cannot be included between the observed planet pairs without destabilizing the system, indicative of previous dynamical sculpting \citep{Barnes&Quinn2004, Fang&Margot2013, Obertas2023}. Our conclusion --- that intra-system uniformity can emerge naturally from dynamical sculpting --- provides further support for the importance of a giant impact phase in shaping the architectures of exoplanet systems.

On its face, establishing intra-system uniformity via giant impacts appears to present a challenge for the retention of the gaseous envelopes found to cover super-Earths/sub-Neptunes, since giant impacts are believed to strip the H/He atmospheres of such planets \citep{Inamdar&Schlichting2015, Biersteker&Schlichting2019, Poon2020}. However, substantial gaseous envelopes can still be accreted after the dynamical influence of the gas disk has waned, because accretion onto small planets is regulated by the cooling rate of the envelope rather than the supply of gas \citep{Lee&Chiang2015}. Indeed, \citet{Lee&Chiang2016} propose that small planets must accrete their envelopes only after the gas density is significantly depleted because otherwise they would experience runaway accretion leading to the formation of Jupiter-mass planets. Thus, our results are suggestive of a scenario in which the final assembly of small planet cores is forestalled until the gas disk has mostly dispersed, at which point dynamical instabilities lead to systems comprised of uniformly-spaced, similar-mass planets. The planets can then accrete similar-sized envelopes from the residual gas disk, establishing radius uniformity.

We note that the expected dispersal timescale of gas disks is somewhat uncertain (photoevaporative dispersal can take place in as short as ${\sim}\,10^5$\,years; \citealt{Alexander2006}), but the time required is likely longer than the nominal time at which the first collisions occur in our simulations (${\sim}\,10^6\,P_1$ or $\,{\sim}\,0.03$\,Myr; see Fig.~\ref{fig:trackingD}). As such, in this scenario, giant impacts may begin to take place before the gas disk has fully dispersed. We defer more careful study of this residual gas phase to future work.

This paper presents a re-interpretation of the peas-in-a-pod trend, arguing that intra-system uniformity likely arose as a result of dynamical sculpting rather than the early planet formation process. More broadly, our results provide an example of subtle effects emerging from dynamical sculpting and caution against drawing conclusions about the planet formation process without careful consideration of possible dynamical effects.

\section*{Acknowledgements}

We thank the anonymous referee for their valuable comments. We would also like to thank Sarah Millholland, Lauren Weiss, Max Goldberg, and Konstantin Batygin for useful feedback and discussions. This work was supported in part by the Natural Sciences and Engineering Council of Canada. This research has made use of the NASA Exoplanet Archive, which is operated by the California Institute of Technology, under contract with the National Aeronautics and Space Administration under the Exoplanet Exploration Program.

\section*{Data availability}

The data underlying this article will be shared on reasonable request to the corresponding author.

\bibliographystyle{mnras}
\bibliography{refs}

\end{document}